\documentclass[letterpaper,preprint,superscriptaddress,floatfix]{revtex4}
\usepackage[latin1]{inputenc}
\usepackage{bm}
\usepackage{multirow,amssymb,amsbsy,amsmath}
\usepackage{graphicx}
\usepackage{verbatim}
\usepackage[dvipdfx,cmyk]{xcolor}
\makeatletter
\usepackage{pifont}
\makeatother

\newcommand{\ket}[1]{\ensuremath{\left|#1\right\rangle}}

\begin{document}

\title{
Achieving Heisenberg-scaling precision with
projective measurement on single photons
 }

\author{Geng Chen}
\affiliation{CAS Key Laboratory of Quantum Information, University of
Science and Technology of China, Hefei, 230026, China}
\affiliation{Synergetic Innovation Center of Quantum Information and Quantum Physics, University of Science and Technology of China, Hefei, Anhui 230026, China}

\author{Lijian Zhang$\footnote{email:lijian.zhang@nju.edu.cn}$}
\affiliation{National Laboratory of Solid State Microstructures and College of Engineering and Applied Sciences, Nanjing University, Nanjing, China}
\affiliation{Collaborative Innovation Center of Advanced Microstructures, Nanjing University, Nanjing 210093, China}

\author{Wen-Hao Zhang}
\affiliation{CAS Key Laboratory of Quantum Information, University of
Science and Technology of China, Hefei, 230026, China}
\affiliation{Synergetic Innovation Center of Quantum Information and Quantum Physics, University of Science and Technology of China, Hefei, Anhui 230026, China}

\author{Xing-Xiang Peng}
\affiliation{CAS Key Laboratory of Quantum Information, University of
Science and Technology of China, Hefei, 230026, China}
\affiliation{Synergetic Innovation Center of Quantum Information and Quantum Physics, University of Science and Technology of China, Hefei, Anhui 230026, China}

\author{Liang Xu}
\affiliation{National Laboratory of Solid State Microstructures and College of Engineering and Applied Sciences, Nanjing University, Nanjing, China}
\affiliation{Collaborative Innovation Center of Advanced Microstructures, Nanjing University, Nanjing 210093, China}

\author{Zhao-Di Liu}
\affiliation{CAS Key Laboratory of Quantum Information, University of
Science and Technology of China, Hefei, 230026, China}
\affiliation{Synergetic Innovation Center of Quantum Information and Quantum Physics, University of Science and Technology of China, Hefei, Anhui 230026, China}

\author{Xiao-Ye Xu}
\affiliation{CAS Key Laboratory of Quantum Information, University of
Science and Technology of China, Hefei, 230026, China}
\affiliation{Synergetic Innovation Center of Quantum Information and Quantum Physics, University of Science and Technology of China, Hefei, Anhui 230026, China}

\author{Jian-Shun Tang}
\affiliation{CAS Key Laboratory of Quantum Information, University of
Science and Technology of China, Hefei, 230026, China}
\affiliation{Synergetic Innovation Center of Quantum Information and Quantum Physics, University of Science and Technology of China, Hefei, Anhui 230026, China}

\author{Yong-Nan Sun}
\affiliation{CAS Key Laboratory of Quantum Information, University of
Science and Technology of China, Hefei, 230026, China}
\affiliation{Synergetic Innovation Center of Quantum Information and Quantum Physics, University of Science and Technology of China, Hefei, Anhui 230026, China}

\author{De-Yong He}
\affiliation{CAS Key Laboratory of Quantum Information, University of
Science and Technology of China, Hefei, 230026, China}
\affiliation{Synergetic Innovation Center of Quantum Information and Quantum Physics, University of Science and Technology of China, Hefei, Anhui 230026, China}

\author{Jin-Shi Xu}
\affiliation{CAS Key Laboratory of Quantum Information, University of
Science and Technology of China, Hefei, 230026, China}
\affiliation{Synergetic Innovation Center of Quantum Information and Quantum Physics, University of Science and Technology of China, Hefei, Anhui 230026, China}

\author{Zong-Quan Zhou}
\affiliation{CAS Key Laboratory of Quantum Information, University of
Science and Technology of China, Hefei, 230026, China}
\affiliation{Synergetic Innovation Center of Quantum Information and Quantum Physics, University of Science and Technology of China, Hefei, Anhui 230026, China}

\author{Chuan-Feng Li$\footnote{email:cfli@ustc.edu.cn}$}
\affiliation{CAS Key Laboratory of Quantum Information, University of
Science and Technology of China, Hefei, 230026, China}
\affiliation{Synergetic Innovation Center of Quantum Information and Quantum Physics, University of Science and Technology of China, Hefei, Anhui 230026, China}

\author{Guang-Can Guo}
\affiliation{CAS Key Laboratory of Quantum Information, University of
Science and Technology of China, Hefei, 230026, China}
\affiliation{Synergetic Innovation Center of Quantum Information and Quantum Physics, University of Science and Technology of China, Hefei, Anhui 230026, China}

\begin{abstract}
It has been suggested that both quantum superpositions and nonlinear interactions are important resources for quantum metrology. However, to date the different roles that these two resources play in the precision enhancement are not well understood. Here, we experimentally demonstrate a Heisenberg-scaling metrology to measure the parameter governing the nonlinear coupling between two different optical modes. The intense mode with $n$ (more than $10^6$ in our work) photons manifests its effect through the nonlinear interaction strength which is proportional to its average photon-number. The superposition state of the weak mode, which contains only a single photon, is responsible for both the linear Hamiltonian and the scaling of the measurement precision. By properly preparing the initial state of single photon and making projective photon-counting measurement, the extracted classical Fisher information (FI) can saturate the quantum FI embedded in the combined state after coupling, which is $\sim n^2$ and leads to a practical precision $\simeq 1.2/n$. Free from the utilization of entanglement, our work paves a way to realize Heisenberg-scaling precision when only a linear Hamiltonian is involved.
\end{abstract}
\date{\today}

\maketitle

\section{Introduction}
%\LZ{we should cite DQC1 metrology, Jordan and Aharonov's paper.}

Quantum metrology promises the enhancement of measurement precision beyond the limit of classical methods, therefore have received substantial interest due to its potential scientific and technical applications \cite{Giovannetti,Helstrom,Holevo,Yuan,Wineland,Caves,Lee,Braunstein,Giovannetti1,Tan,Pirandola,Nair}.  %A lot of theoretical and experimental efforts have been devoted to this subject recently \cite{Giovannetti,Helstrom,Holevo,Yuan,Wineland,Caves,Lee,Braunstein,Giovannetti1,Tan,Pirandola,Nair}.
For the estimation of a parameter $g$ with a meter state that contains on average $n$ particles, a major way to achieve the quantum-enhanced precision is by making use of entanglement \cite{Bollinger,Walther,Afek}, resulting in an improved precision that surpass the standard quantum limit (SQL) which scales as $\delta g \propto 1/\sqrt{n}$ or even achieve the Hisenberg scaling, $\delta g \propto 1/n$. However, the difficulty to fabricate large-scale entangled states and the fragility of such states make it challenging for quantum-enhanced schemes to surpass classical techniques in practical applications. On the other hand whether entanglement is a necessary resource for quantum-enhanced precision is under debate \cite{Hyllus,Tilma}. Indeed, schemes exploiting non-classicality and quantum coherence other than entanglement \cite{Braun}, for example nonlinear Kerr effect \cite{Chen}, quadrature squeezing \cite{Goda,Grangier,Xiao,Treps} and sequential measurement with single-particle superposition states \cite{Higgins,Waldherr}, have achieved scaling beyond $1/\sqrt{n}$ as well. These results motivate the quest for novel precision metrology schemes.

%Recently it has been shown that quantum metrology with Hamiltonians involving interactions among particles can achieve the precision beyond the $1/\sqrt{N}$-scaling without non-classical states (\LZ{citations}), rendering the nonlinear interaction another useful resource for precision measurement. The applications of nonlinearity in quantum metrology also refine the definition of SQL and HL (\LZ{cite PRL 105, 180402 (2010)}). In particular wether non-classical states still perform better than classical states in nonlinear quantum metrology is unclear (\LZ{cite PRA 80, 063811 (2009) and PRA PRA 81, 022108}).On the other hand, recently there have been increased efforts to apply the weak-value amplification (WVA) in weak measurement \cite{Aharonov,Aharonov1,Steinberg,Hosten,Dixon} to precision metrology \cite{Brunner3,Xu,Lyons,Pang,Pang1}. However, it is widely acknowledged that weak measurements cannot effectively improve signal to noise ratio or measurement sensitivity when the probability decrease due to postselection needs to be considered \cite{Zhang,Vaidman,Zhu}. Especially, to date almost all of the WVA measurements are performed in the configuration space, and therefore have a maximum precision at the SQL.

 Here we propose and experimentally demonstrate a quantum scheme which can be viewed as a projective photon-counting measurement (PPCM) in the phase space. Our scheme relies on the coupling between a single-photon superposition state and an intense coherent beam with an average photon-number $n$, and achieves a $1/n$ scaling, the Heisenberg scaling (HS), in the measurement precision up to $n=5\times 10^6$. Our proposal seems similar to the so-called weak-value amplification (WVA) technique \cite{Aharonov,Aharonov1,Steinberg,Hosten,Dixon,Brunner3,Xu,Lyons,Pang,Pang1} by projecting the single photon state onto a nearly orthogonal basis. However, it is fundamentally different from the previous WVA measurement: instead of relying on weak-value amplification in the post-selected meter state \cite{Zhang,Vaidman,Zhu}, we use the projective probabilities to extract the information. Post-selection measurement schemes generally decrease signal to noise ratio(because they involve throwing away data) and that measuring the whole signal (as is done here) will generally give better precision.

%The other way is to utilize the magnification effect with weak measurement, which shows particular advantageous in weak coupling regime \cite{Aharonov,Aharonov1,Steinberg,Hosten,Dixon}. This magnification effect is at the cost of a reduced rate at which data can be acquired due to the requirement of post-selection. As a result, weak measurement cannot effectively improve the signal to noise ratio and the measurement sensitivity when the probability decrease due to post-selection needs to be considered \cite{Zhang,Ferrie}.

%Many technical innovations have been proposed to standard weak measurement aiming to improving the precision \cite{Brunner3,Xu,Lyons,Pang,Pang1}. One adoptable way is to retain all post-selection results rather than retaining only one specific result \cite{Dressel,Wang}. However, all these proposals still attempt to measure the shift on meter states and the resulted scaling still follows SQL. In Ref. \cite{Zhang}, Zhang $\emph{et al.}$ divides the extractable Fisher information (FI) \cite{Fisher} of a standard weak measurement process into three parts. It is proved that besides retaining only the information in successfully post-selected meter states, the failed part and post-selection itself also provide inevitable amount of FI. In particular, when the investigated Hamiltonian describes a phase-space interactions, the post-selection process has much more information and indeed scales at the Heisenberg limit.

The PPCM makes the scheme robust and easy to implement, more important, a nearly optimal HS precision can be attained for the measurement of single photon Kerr nonlinearity in a photonic crystal fibre (PCF) \cite{Russell}. In experiment, we observe an ultra small Kerr nonlinearity of $\simeq 6 \times 10^{-8}$ rad. The practical HS achieved here can be expressed as $\simeq 1.2/n$, with the best precision to be $\simeq 1 \times 10^{-10}$ rad when $n=5\times 10^6$. This result significantly improves previous similar tasks, both on the scaling and best precision achieved.

\section{Underlying theoretical analysis}
\begin{figure}
\centering \includegraphics[width=6in]{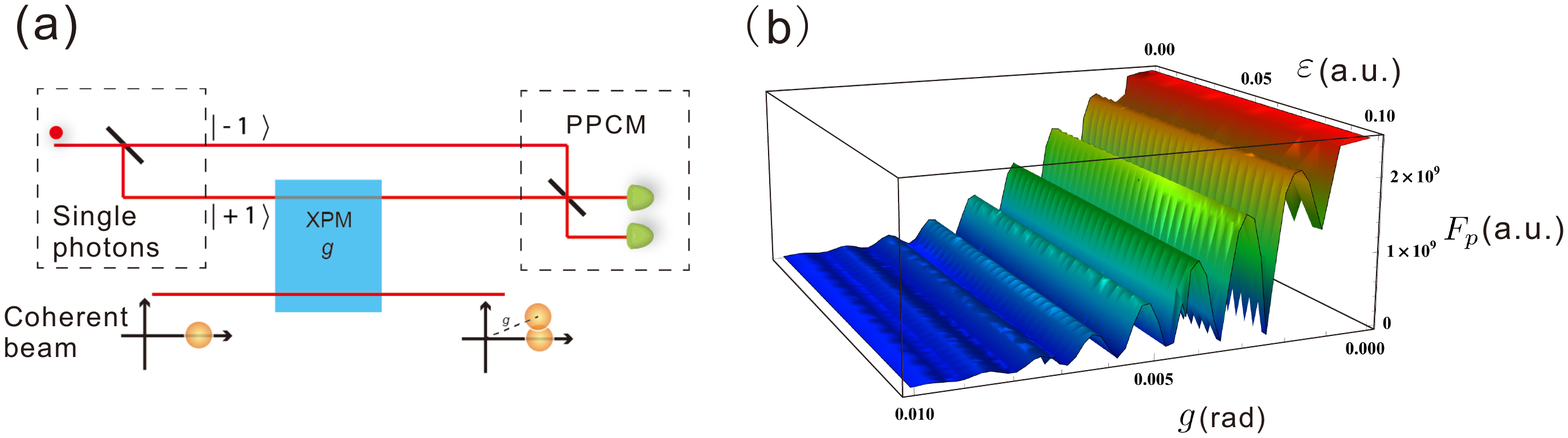}
\caption{(a) The scheme: a single-photon goes through an interferometer, where in one arm it interacts with a strong laser pulse through the XPM and a phase is acquired, while in the other nothing happens. At the exit port the two paths interfere so the probability of the photon coming out there depends on the phase it acquired by the interaction, and thus, also on the phase of the strong pulses. A standard method to measure the phase of the strong pulses merely results in a precision bounded by SQL, while performing PPCM on singe photons one can observe an HS precision. (b) The classical information $F_{p}$ extracted by PPCM for varying interaction strength $g$ and projective parameter $\varepsilon$, when the mean photon number of a strong pulse is $n=5\times10^{4}$. As $g\rightarrow0$, $F_{p}$ becomes dominant in $F_{tot}$ and scales in $n^{2}$, which means a practical HS precision is attainable by measuring the accepted and rejected probabilities of PPCM. }
\label{FIplot}
\end{figure}

Considering the practical task to measure the parameter governing the nonlinear coupling, i.e. cross phase modulation (XPM) effect, between two different optical modes \cite{Zhang}, which are single photons and strong pulses from coherent beam, respectively, as shown in Fig. \ref{FIplot}(a). The single photons are in superposition state between the two arms of the interferometer, which can be
written as
\begin{equation}
\label{Prestate}
|\psi_{i}\rangle = \cos(\theta_{i}/2)|U\rangle + \sin(\theta_{i}/2)|D\rangle,
\end{equation}
where $|U\rangle$ and $|D\rangle$ denote the wavepacket in the upper and down arms of the interferometer in Fig. 1(a).
In view of the fact that only the $|D\rangle$ wavepacket can interact with the coherent beam, Eq. (\ref{Prestate}) can be written to be the superposition of photon number states joining the interaction, as
\begin{equation}
\label{Photonnumber}
|\psi_{i}\rangle = \cos(\theta_{i}/2)|0\rangle_{U} + \sin(\theta_{i}/2)|1\rangle_{D},
\end{equation}
where $|0\rangle$ and $|1\rangle$ denote the photon number interacting with coherent beam.
 The initial
state of the strong pulses is coherent state $|\alpha\rangle$. The coupling
strength $\emph{g}$ is the parameter to be estimated, it appears in the Hamiltonian $H = -g\delta(t-t_{0})\hat{n_{0}}\hat{n}$, where $\hat{n_{0}}$ and $\hat{n}$ are the particle number operators for weak optical mode and strong pulse part, respectively. \textcolor[rgb]{1.00,0.50,0.00}{In this interacting Hamiltonian, the parameter in question $g$
multiplies a Hamiltonian term that multiplicatively couples two modes,
not just an operator on a single mode. Especially, when the weak optical mode is single photons, the Hamiltonian reduces to a linear form with respect to $n$. Consequently, $n$ can be defined as the photon number used in each measurement.}

After this interaction, the
joint state becomes

\begin{equation}
\label{Joint}
|\psi_{i}\rangle = \cos(\theta_{i}/2)|0\rangle_{U}|\alpha\rangle+ \sin(\theta_{i}/2)|1\rangle_{D}|\alpha e^{ig}\rangle.
\end{equation}

\textcolor[rgb]{1.00,0.50,0.00}{This joint state contains an amount of quantum FI as} \cite{Zhang}

\begin{equation}
\label{QFI}
Q_{j}=n^{2}\sin^{2}\theta_{i}+n[\sin^{2}(\theta_{i}/2)]
\end{equation}
where $n = \mid\alpha\mid^{2}$ is the
mean photon number of the strong pulses. \textcolor[rgb]{1.00,0.50,0.00}{The $n^{2}$ scaling of $Q_{j}$ suggests that an HS can be achieved with optimal measurement strategy, which is similar to the precision-enhancement effect of multi-qubit entangled state \cite{Giovannetti1}. A distinct advantage is that this joint state is much more straightforward to produce than an entangled state of multi-qubits}.

The Cram\'{e}r-Rao bound, which is defined as
$\Delta^{2}g \geq 1/(\nu F_{tot})$ \cite{Braunstein1}, decides the best achievable precision $\Delta g$
in estimating $g$. $F_{tot}$ is the sum total of the
classical and quantum FI contained
at different stages of the PPCM process and $\nu$ is the times of performing PPCM.  The quantum FI $Q_{j}$ is the upper bound of $F_{tot}$ because the PPCM is a special case of the global measurement
on the joint state in Eq. (\ref{Joint}).

Projecting the single photons into state $|\psi_{f}\rangle = \cos(\theta_{f}/2)|U\rangle + e^{i(\pi-\varepsilon)}\sin(\theta_{f}/2)|D\rangle$, the probabilities of the accepted and rejected events are denoted as $P_{d}$ and $P_{r}$, respectively.
Afterwards, FI can be divided into three parts and we have $F_{tot}=P_{d}Q_{d}+P_{r}Q_{r}+F_{p}$. $Q_{d}$ and $Q_{r}$ denote the quantum FI contained in strong pulses for accepted and rejected cases, respectively. $F_{p}$ is the FI in the PPCM process itself, which can be exactly calculated as

\begin{equation}
\label{Fp}
F_{p}=\frac{n^{2}(\sin\theta_{i}\sin\theta_{f}exp(-2n\sin^{2}g))^{2}\sin^{2}(n\sin2g+2g+\pi-\varepsilon)}{1-[\sin\theta_{i}\sin\theta_{f}exp(-2n\sin^{2}g)
\cos(n\sin2g+\pi-\varepsilon)-\cos(\theta_{i})\cos(\theta_{f})]^{2}}.
\end{equation}

Due to the decay factor $exp(-2n\sin^{2}g)$ in the numerator, $F_{p}$ decreases very fast with increasing $g$ for a given value of $n$. As it is shown in Fig. 1(b) for $n=5\times10^{4}$, when the interaction strength $g$ is extremely small, $F_{p}$ approximately equals to $n^{2}$ for all plotted $\varepsilon$. As $g$ increases, a rapid oscillation decay can be observed and eventually $F_{p}$ approaches 0 when $g$ exceeds 0.01.

When $\theta_{i}=\theta_{f}=\pi/2$ and $g\rightarrow0$, we can get

\begin{align}
& F_{p}=n^{2} \\
& P_{d}Q_{d}=(1-\varepsilon^{2}/4)n \\
& P_{r}Q_{r}=\varepsilon^{2}n/4
\end{align}
and the accepted probability is
\begin{equation} \label{Pd}
P_{d}=\frac{1-\cos(2gn+\varepsilon)}{2}.
\end{equation}
These results show that quantum-enhanced scaling can
be attained by performing a PPCM on the single photons. This is the primary change compared to earlier experiments, \textcolor[rgb]{1.00,0.50,0.00}{which concentrate on the measurement of strong mode} \cite{Chen,Steinberg}. Yet, as indicated by Eqs. (6)-(8), the PPCM process has much more information for large $n$ and small $g$, and indeed scales at the HS.

\section{Experiment results}

\begin{figure}
\centering \includegraphics[trim=1cm 0.8cm 2.0cm 0.5cm, clip=true,width=0.7\textwidth]{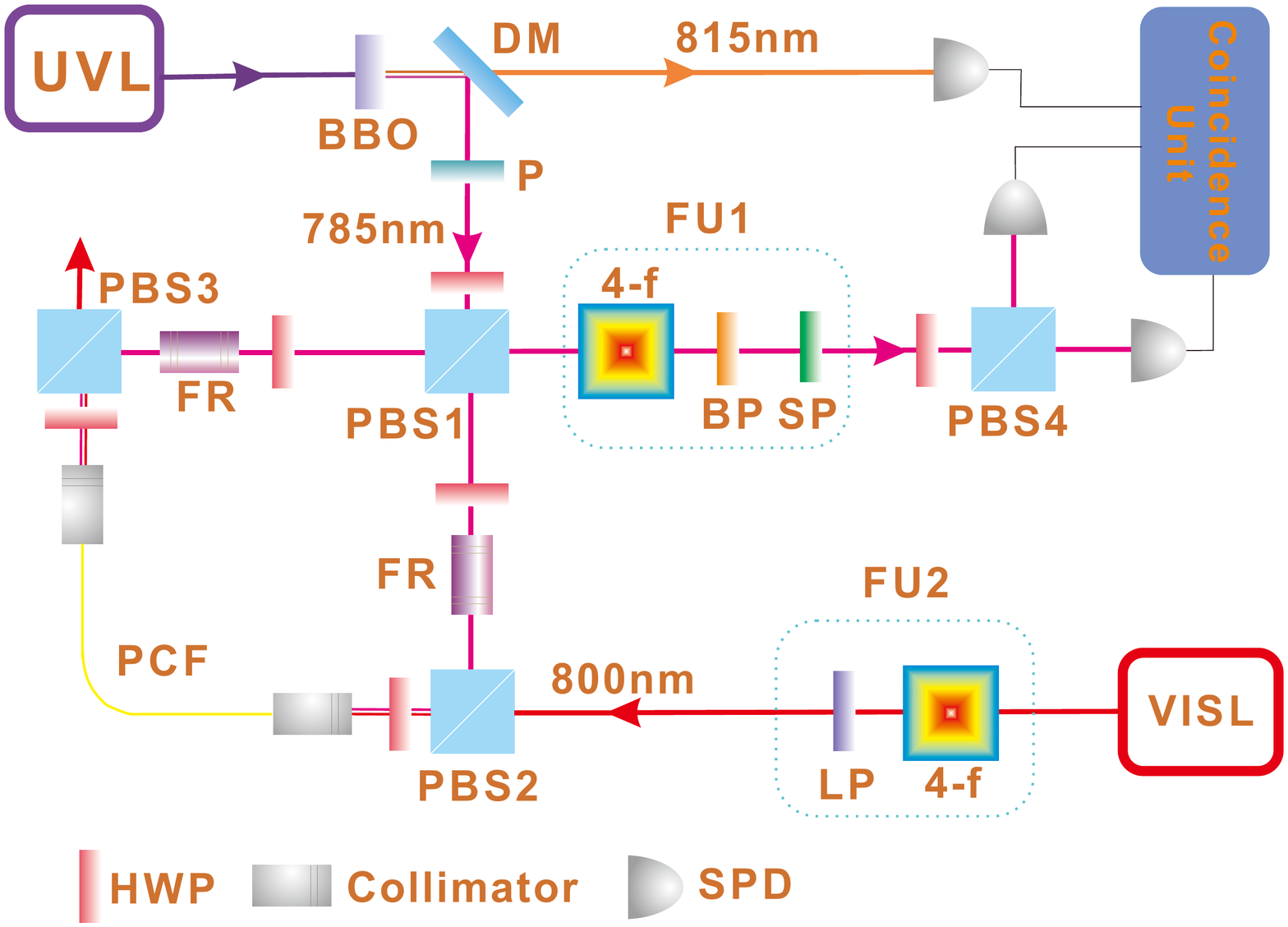}
\caption{\textbf{Experimental setup}:
 The 815 nm photons serve as triggers and the heralded 785 nm photons interact with strong pulses (800 nm) in an 8 m long photonic crystal fiber (PCF), centering in a PSI. The interaction strength $g$ is estimated from the distribution of the accepted and rejected events of PPCM.
 BBO: $\beta$-barium borate crystal; DM: dichroic mirror; SPD: single-photon detector; HWP: half wave plate; P: polarizer; FR: Faraday rotator; 4-f: 4-f filtering unit; PBS: polarized beam splitter; BP: band pass filter; LP: long-wavelength-pass filter; SP: short wavelength pass filter; UVL: ultra-violet laser; VISL: visible laser.
}\label{setup}
\end{figure}

The photon-interaction scenario can be experimentally investigated with the setup shown in Fig. \ref{setup}. The two interacting parties are heralded single photons and strong pulses in coherent states. Pumping the $\beta$-barium borate (BBO) crystal by a UV laser (UVL), the single photons can be generated through a non-degenerate spontaneous parametric down conversion (SPDC) process. After separating from the 815 nm heralding photons, the 785 nm single photons is prepared as $\psi=(|H\rangle+|V\rangle)/\sqrt{2}$ ($\emph{V}$ and $\emph{H}$ represent the vertical and horizontal polarization respectively). When entering the polarization Sagnac interferometer (PSI), the photon is in an equal superposition of clockwise and counter-clockwise propagation. The PSI mainly contains three PBSes, two Faraday units and an 8 m long photonic crystal fiber (NL-2.4-800, Blaze Photonics). PBS1 acts as the entrance and exit ports of PSI, from which the 785 nm single photons enters and leaves PSI. The horizontally polarized strong pulses from
an 800 nm visible laser (VISL) are coupled into PSI by PBS2. The strong pulses are synchronized and collinear with 785 nm single photons, afterwards they are coupled into PCF by two triplet fiber optic collimators (Thorlabs TC12FC-780) with a coupling efficiency of 20$\%$. Only the clockwise component of single photons can interact with the strong pulse; hence, the photon number state becomes $\ket{\psi} = (\ket{1}_{H} + \ket{0}_{V})/\sqrt{2}$, where $\{0,1\}$ represents the interacting photon number. With an HWP before each collimator, photon polarization is maintained after the PCF. Two Faraday units, each consisting of a 45$^{\circ}$ Faraday rotator and an HWP, cause the two counter-propagating components to have the same linear polarization in the PCF. After exiting the PCF, strong pulses depart the PSI from PBS3 and two counter-propagating components of single photons are finally combined at PBS1. After leaving PSI from PBS1, single photons are projected into $|\psi_{f}\rangle = |H\rangle + e^{i(\pi-\varepsilon)}|V\rangle$, by an HWP followed by PBS4. The rates of accepted and rejected PPCM events are simultaneously recorded by a multi-channel coincidence unit.

\begin{figure}
\centering
\includegraphics[width=5.5in]{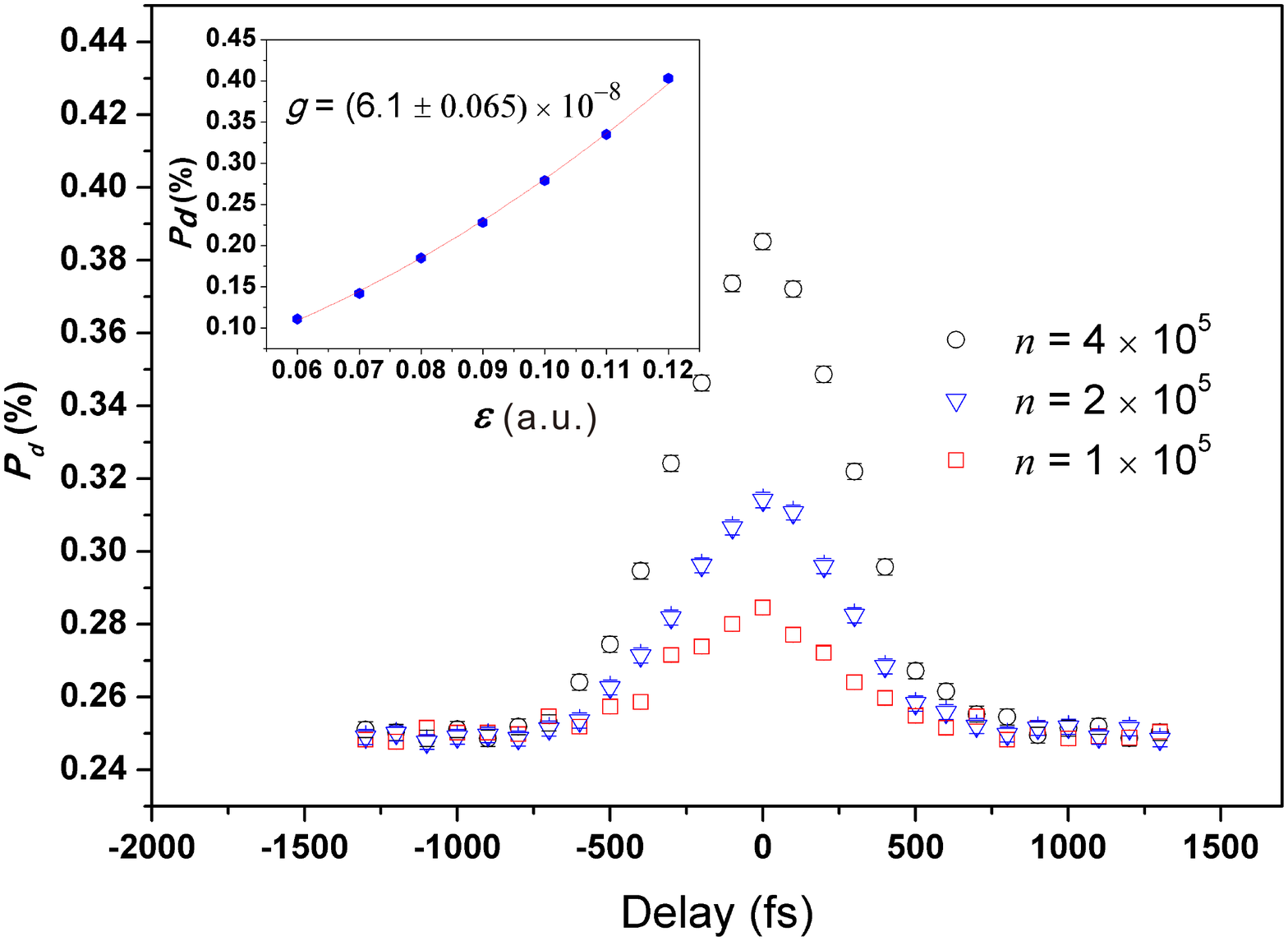}
\caption{(a) Changing the time-domain overlap of single photons and strong pulses, the interaction strength $g$ can be adjusted and $P_{d}$ varies dependently. The largest $P_{d}$ is obtained when $g$ reaches a maximum at zero delay point. The inset shows the dependence of $P_{d}$ on the projective parameter $\varepsilon$ with $n=1\times10^{5}$. The measured $P_{d}$ can be well fitted with Eq. (\ref{Pd}) giving $g$ to be $(6.1 \pm 0.065)\times10^{-8}$ rad.}\label{delay}
\end{figure}

The measured accepted probability of PPCM can be calculated as $P_{d}=N_{d}/(N_{d}+N_{r})$, where $N_{d}$ and $N_{r}$ represent the total number of accepted and rejected events of PPCM. First of all, in order to obtain the maximal interaction strength $g_{0}$, we measure $P_{d}$ while scanning the time-domain delay between single photons and strong pulses, as shown in Fig. \ref{delay}. The strong pulses, which are centering at the zero-dispersion point of PCF, come from femtosecond lasers with a full wave at half maximum of 150 fs. Refer to the single photon, it should be broadened due to the dispersion in the PCF. The shape of the single photons can be inferred from the time-delay measurement in Fig. \ref{delay} by fitting the data with Gaussian function, and we can conclude that the single photons are also approximately Gaussian-shaped with a full wave at half maximum of $\simeq 480$ fs. Considering these parameters, we can calculate the maximum overlap at zero delay to be 0.77. Three values of $n$ (photon number of strong pulses) are investigated when $\varepsilon=0.1$. For all of them $P_{d}$ reaches the maximum at an identical delay point, when the two interacting parties overlap completely in the PCF. Fixed at this point, we measure $P_{d}$ for seven $\varepsilon$ values and a fitting analysis with Eq. (\ref{Pd}) gives that $g_{0}$ = $(6.1 \pm 0.065)\times10^{-8}$ rad. Therefore, changing the time domain overlap enables us to adjust $g$ from 0 to $g_{0}$ continuously. The calibration method here is also a PPCM, which is completely different from the weak-value amplification technique in Ref. \cite{Chen}.

\begin{figure}

\centering \includegraphics[trim=0cm 0cm 2.0cm 0cm, clip=true,width=0.75\textwidth]{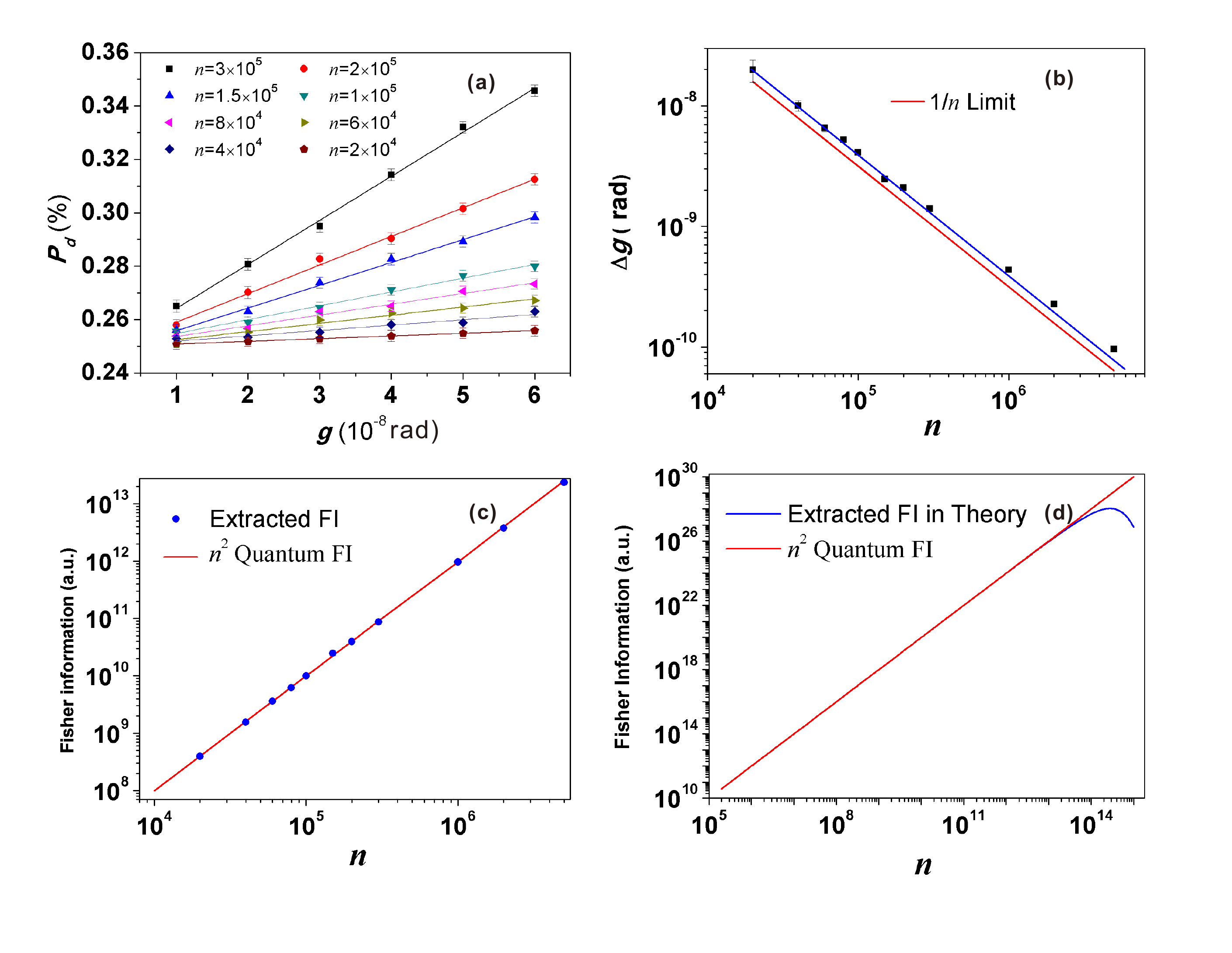}
\caption{(a) For each value of $g$, $P_{d}$ is measured for a number of values of $n$ and the measurement sensitivity $s$ is defined as the linear fitting slope of the solid lines. The uncertainty $\delta P$ is selected to be the standard error of the measurement results. The ratio of $\delta P$ and $s$ characterizes the overall precision $\Delta g$. (b) The precision $\Delta g$ for a number of values of $n$ from $2\times10^{4}$ to $5\times10^{6}$.
The blue line is the linear fitting of the data points representing an HS as $1.2/n$, while the red line stands for $1/n$ limit. (c) The amount of extracted classical FI for different $n$. The $n^{2}$ scaling indicates $1/n$ limit is allowed in this measurement. (d) The theoretical simulation of extracted FI when $n$ is further increased to $10^{15}$. The scaling deviates away from $n^{2}$ when $n$ is larger than $10^{13}$.}\label{HL}
\end{figure}

It has been shown that $g$ can be precisely estimated directly from $P_{d}$. Both the FI calculation and the signal-noise analysis reveal that this method is adequate to reach HS precision. Nonetheless, it is worthwhile to verify that the HL can be experimentally attained with realistic conditions. In practice, the precision $\Delta g$ of this measurement can be estimated as $\Delta g$ = $\delta P$ / $s$,
where $s={\partial P_{d} \over \partial g}$ stands for the sensitivity of this method and the uncertainty $\delta P$ can be calculated as \cite{John}

\begin{equation}
\label{uncertainty}
\delta P=\sqrt{\sum_{i=1}^{\nu}(\frac{\partial P_{di}}{\partial P_{d}}\sigma)^2}=\sigma / \sqrt{\nu},
\end{equation}
where $P_{d}$ and $\sigma$ are the mean and standard deviation of $\nu$ times measurement results, which are denoted as $P_{di}$ (i=1,2...,$\nu$). In order to determine $s$, $P_{d}$ is measured for a series of $g$ values. Tuning the time domain overlap of single photons and strong pulses, each $g$ value is well calibrated from Eq. (\ref{Pd}) with $n = 6 \times 10^{5}$, $\varepsilon = 0.1$, when $N_{d}+N_{r} \simeq 5 \times 10^{7}$. Afterwards, we measure $P_{d}$ with different $n$ for a certain $g$ value, as shown in the inset of Fig. \ref{HL}(a). For each data point, we perform $\nu$ = 10 times measurement and each time we record about 110 seconds with $N_{d}+N_{r}$ = $1\times10^{6}$. Making statistical analysis on these results, $P_{d}$ and $\sigma$ are selected as the mean and standard deviation, respectively. Changing $g$ from $1\times10^{-8}$ to $6\times10^{-8}$ rad we can estimate $s$ from the linear fitting slope. Because $s$ is approximately proportional to $n$ and the uncertainty is roughly a constant, the overall precision is inversely proportional to $n$, as shown in Fig. \ref{HL}(b). As a result, the HS precision is experimentally verified and the ultimate precision we get is $1\times10^{-10}$ rad.
In principle, further increasing $n$ the HS maintains and we can get better precision. However, in experiment, the noise photons leaking from 800 nm strong pulses will also increase to damage the precision. When $n$ is larger than $10^{6}$, the leaking 800 nm photons are no longer negligible and the measured precision slightly deviates from the HS as shown in Fig. 4(b).

With the measured $P_{d}$ and $s$, the extracted classical FI can be calculated as \cite{Zhang}
\begin{equation}
\label{FI}
F_{p}=\frac{1}{P_{d}}(\frac{\partial P_{d}}{\partial g})^{2}+\frac{1}{1-P_{d}}(\frac{\partial (1-P_{d})}{\partial g})^{2},
\end{equation}
and the results are shown in Fig. \ref{HL}(c). The $n^{2}$ scaling indicates that our method can in principle attain the $1/n$ limit, simply with a PPCM on single photons. Practically, \textcolor[rgb]{1.00,0.50,0.00}{technical} noises in current experiment can damage the precision, and thus, the obtained scaling is slightly worse than $1/n$ limit, as shown in Fig. 4(b). The uncertainty in $P_{d}$ from shot noise is $\sim10^{-5}$ for each point in Fig. 4(a), however, kinds of \textcolor[rgb]{1.00,0.50,0.00}{technical} noises can add to this uncertainty and damage the HS (see Supplemental Material \cite{Supp} for details).

\section{Discussion}

In this experiment, we investigate the XPM effect \cite{Matsuda2008} between a single photon and a strong pulse. The interaction strength here only depends linearly on $n$, in contrast to previous nonlinear metrology utilizing self-phase modulation (SPM)\cite{Napolitano,Dur,Hall}. The additional noise introduced by SPM is a main limit in these nonlinear metrology. However, the SPM is insignificant here since the strong pulses are discarded and only the single photons are measured finally.

It may not be surprising that in the presence of photon interactions the quantum FI shows a quantum-enhanced scaling of $\sim n^{2}$ \cite{Zhang}. However, it is still a main challenge to extract a large amount of the information, namely classical FI, and eventually achieve a practical HS.  In previous similar tasks, the strong pulses which contain substantial amount of photons are measured \cite{Matsuda2008,Steinberg} and the precision is standard-quantum-limited to estimate $g$. Recently work shows that measuring the photon number shift in a mixed probe \cite{Chen}, an HS is observed as $3/n$ when $n<10^{5}$, corresponding to an FI as $\simeq 0.11n^{2}$. In this work, in spite of containing only a single photon, the weak mode is responsible for the final HS as $1.2/n$ with the FI to be $\simeq 0.69n^{2}$. Contrastively, strong pulses containing $\sim10^{6}$ photons only provide a negligible amount of FI and are completely discarded after interacting with single photons. A further theoretical simulation reveals that this HS maintains when $n<10^{13}$, as shown in Fig. 4(d). Therefore, it can be concluded that this method is much more accurate and robust than previous works.

The HS precision here seems to be the result of a simple signal amplification effect by increasing $n$, however, this is not the source of precision enhancement. Undoubtedly, purely increasing $n$ (the mean photon number of strong pulses) we can get larger signal. On the other hand, the error also rises due to the quantum fluctuation in coherent state itself. A trival measurement, including standard interferometers \cite{Matsuda2008} and standard weak measurement \cite{Steinberg}, cannot beat standard quantum limit and the precision still scales as $1/\sqrt{n}$. The situation changes if we measure the probabilities in the PPCM instead of the phase (or energy) shift in the strong pulses. The signal in our method is approximately proportional to $n$ while the error is nearly a constant with increasing $n$.

 In this work, using coherent states with the mean photon number up to $\sim 10^{6}$, we implement a practical measurement task and show that the HS can be attained by simply recording the accepted and rejected probabilities in PPCM. This classical statistical information results in a precision of $\simeq 1\times10^{-10}$ rad when measuring the Kerr nonlinearity of single photon level. Similar schemes may also find applications in cavity or circuit quantum electrodynamics systems \cite{Jordan}. Our results shed new light on both the understanding of quantum metrology and weak measurement, hence can be instructive for the development of new technologies in practical measurement tasks.

 {\bf  Acknowledgments}

This work was supported by the National Key Research and Development Program of China (Nos. 2017YFA0304100, 2016YFA0302700), National Natural Science Foundation of China (Grant Nos. 61327901, 91536113, 11474159, 61490711,11774335, 91536219), the Key Research Program of Frontier Sciences, CAS (Grant No. QYZDYSSW-SLH003), the Fundamental Research Funds for the Central Universities (Grant No. WK2030020019, WK2470000026, 021314380079), Anhui Initiative in Quantum Information Technologies (AHY020100, AHY060300).
%\clearpage
{xx}

\end{document}